\documentclass[10pt,preprint] {aastex}
\usepackage{amssymb,amsmath,gensymb}
\usepackage{graphicx,epsfig,natbib,color}
\usepackage{lscape}
\usepackage{graphicx}
\usepackage{epsfig}
\usepackage{natbib}

\slugcomment{Submitted to ApJ}

\begin{document}
\title{A Type II Radio Burst without a Coronal Mass Ejection}

\author{W. Su,\altaffilmark{1,2}, X. Cheng,\altaffilmark{1,2}, M. D. Ding,\altaffilmark{1,2}, P. F. Chen,\altaffilmark{1,2}, J.Q. Sun,\altaffilmark{1,2}}

\affil{$^1$ School of Astronomy and Space Science, Nanjing University, Nanjing 210093, China}\email{dmd@nju.edu.cn; xincheng@nju.edu.cn}
\affil{$^2$ Key Laboratory for Modern Astronomy and Astrophysics (Nanjing University), Ministry of Education, Nanjing 210093, China}

\begin{abstract}
Type II radio bursts are thought to be a signature of coronal shocks. In this paper, we analyze a short-lived type II burst that started at 07:40 UT on 2011 February 28. By carefully checking white-light images, we find that the type II radio burst is not accompanied by a coronal mass ejection, only with a C2.4 class flare and narrow jet. However, in the extreme-ultraviolet (EUV) images provided by the Atmospheric Imaging Assembly (AIA) on board the Solar Dynamics Observatory (SDO), we find a wave-like structure that propagated at a speed of $\sim$ 600 km s$^{-1}$ during the burst. The relationship between the type II radio burst and the wave-like structure is in particular explored. For this purpose, we first derive the density distribution under the wave by the differential emission measure (DEM) method, which is used to restrict the empirical density model. We then use the restricted density model to invert the speed of the shock that produces the observed frequency drift rate in the dynamic spectrum. The inverted shock speed is similar to the speed of the wave-like structure. This implies that the wave-like structure is most likely a coronal shock that produces the type II radio burst. We also examine the evolution of the magnetic field in the flare-associated active region and find continuous flux emergence and cancellation taking place near the flare site. Based on these facts, we propose a new mechanism for the formation of the type II radio burst, i.e., the expansion of the strongly-inclined magnetic loops after reconnected with nearby emerging flux acts as a piston to generate the shock wave.
\end{abstract}


\section{Introduction}
Shock waves are an important and ubiquitous phenomenon in astrophysics. They can accelerate electrons \citep{Mann1995, Miteva2007, Schwartz2011} and ions \citep{Thomsen1985, Sckopke1995, Giacalone2005} effectively. In the solar atmosphere, there exist various kinds of shocks, which usually manifest themselves as wave-like structures in images, or are implied in the radio dynamic spectra. In particular, type II radio bursts, usually following the eruption of coronal mass ejections (CMEs), are a kind of plasma emission with a slow frequency drift in the radio dynamic spectrum, and they are generally thought to be a signature of coronal shocks \citep{Wild1950, Zheleznyakov1970}.

It is believed that the majority of interplanetary shocks (within 1 AU) are CME-driven \citep{Cane1987, Reiner2001, Reiner2001stat}. However, the origin of the metric type II radio bursts is still an open question; they can be generated either by CME-driven shocks \citep{Cliver2004, Liu2009, Chen2011, Cho2013} or by flare-caused blast waves \citep{Leblanc2001, Magdalenic2008, Nindos2011}. This is an important topic but hard to explore since the radio spectrum has usually a low or even no spatial resolution. The situation has been greatly improved since the launch of recent space solar instruments. In particular, the Atmospheric Imaging Assembly \citep[AIA;][]{Lemen2012} on board the Solar Dynamics Observatory \citep[SDO;][]{Pesnell2012} can observe the low corona under 0.4R$_{\odot}$, where the type II radio bursts are initially produced with the start frequency of hundreds of MHz. Recently, \citet{Bemporad2010} derived physical parameters of the pre- and post-shock regions using multi-wavelength observations. \citet{Ma2011} discovered a clear case of a CME-driven shock in the low corona by extreme-ultraviolet (EUV) images. For more quantitative studies of coronal shocks, some new methods have been developed. \citet{Kouloumvakos2013} combined the differential emission measure (DEM) method to calculate the compression ratio of a coronal shock, and found that the shock can be driven by both the nose and the flanks of the expanding bubble. \citet{Zucca2014} used multi-wavelength observations to obtain two-dimensional density, magnetic field, and Alfv\'en speed maps, which help reveal the properties of shocks more accurately. Among these recent works, the type II radio bursts were all associated with CMEs. It is now widely accepted that the CMEs are the source of the metric type II radio bursts. However, there still exist a few type II radio bursts without CMEs. \citet{Magdalenic2012} reported a CME-less type II radio burst, and suggested that the burst was likely to be generated by the flare. \citet{Gopalswamy2001} mentioned that the reconnection jets could be also one of the candidate structure to generate a type II radio burst.

In this paper, we report a type II radio burst without an associated CME. We propose a new generation mechanism for the coronal shocks. The paper is organized as follows. Observations are described in Section \ref{sect:data}, data analysis and results are presented in Section \ref{sect:Analysis and Result}. Some discussions are made in Section \ref{sect:discussion}, followed by a conclusion in Section \ref{sect:conclusion}.

\section{Observations}
\label{sect:data}
A type II radio burst was detected on 2011 February 28, following the occurrence of a C2.4-class flare which was located at N24E45 in the NOAA active region 11164. The radio burst was observed by a number of radio telescopes, such as Learmonth and San Vito Solar Observatories. A coronal wave-like structure, which was coincident with the type II burst, was observed in detail by the AIA in EUV passbands. The Large Angle Spectrometric Coronagraph \citep[LASCO;][]{Brueckner1995} on board the the Solar and Heliospheric Observatory (SOHO) and the inner coronagraph \citep[COR1;][]{Howard2008} on board the Solar Terrestrial Relations Observatory \citep[STEREO;][]{Kaiser2008} provide the white light images, which are used to inspect whether the radio burst is associated with a CME.

In this work, we use the radio data recorded by the Learmonth Observatory, because the station was at day time during the burst. Compared to the data from San Vito, the dynamic spectrum of Learmonth is clearer. The temporal resolution of the dynamic spectrum is 3 s. As shown in Figure 1, the type II radio burst started at 07:40 UT in the fundamental frequency with the start frequency of 109 MHz, and ended at 07:44 UT with the end frequency of about 52 MHz. The linear frequency drift rate is 0.2375 MHz s$^{-1}$ for the fundamental frequency. The burst is obvious in both fundamental and harmonic frequencies. However, we cannot identify the start frequency in the first harmonic frequency because it is out of observing range. There is a weak indication of band-splitting phenomenon in this burst, presumably implying that the shock was not very strong.

The C2.4-class flare was observed by {\it GOES} a few minutes before the type II radio burst. The soft X-ray flux, which is shown as a red curve in Figure 1, began to rise at 07:34:34 UT, and reached the peak at 07:38:42 UT, 90 s before the type II radio burst. Then the flux declined to the pre-flare level in a few minutes. The rise and decay of the soft X-ray flux are roughly symmetrical. The duration of the flare is less than 10 minutes.

Since type II radio bursts are usually accompanied by CMEs, we check coronal images observed by {\it SOHO}/LASCO and {\it STEREO}/COR1 if this is the case in the present event. The fields of view (FOV) of LASCO and COR1 are located hundreds of Mm above the solar limb; therefore, it takes tens of minutes for a CME to propagate to the visible region. We then check the images of LASCO and COR1 in a period that begins 50 minutes after the peak time of the flare (08:00--08:25 UT) as shown in Figure 2. It is clear that there is no indication of any CME in this time period. Since the active region is close to the solar limb, a CME should be detectable if it exists. For this reason, we can conclude that this flare and type II radio burst are not accompanied by a CME.

We also check the EUV images of the low corona observed by {\it SDO}/AIA whose time interval is 12 s. We find that there is a wave-like structure above the active region in the 171 {\AA}, 193 {\AA} and 211 {\AA} running difference images as indicated by the horizontal arrows in Figure 3 (see online movie for details). The wave-like structure appeared at 07:38 UT and was too weak to be detected after 07:42 UT; it propagated quickly during this period with a velocity of $\sim$ 600 km s$^{-1}$. At almost the same time, a jet (indicated by the vertical arrows in Figure 3) ejected. Moreover, in the running difference images whose time interval is 120 s as shown in Figure 4, we find a group of loops that expand during the propagation of the wave-like structure, and continue to expand after the disappearance of the wave-like structure. The expansion of the loops pushes the wave-like structure to propagate outward. Note that the time interval and contrast ratio of the running difference images are differently adjusted in Figure 3 and Figure 4, in order that they can reveal more clearly the wave-like structure and loops, respectively.

\section{Data Analysis and Result}
\label{sect:Analysis and Result}

From Figures 1 and 3, we can see that the burst and the wave-like structure occured at about 07:40 UT, while the type II radio burst appeared about 2 minutes after the wave-like structure can be initially identified in the AIA images, and it disappeared about 2 minutes after the wave-like structure became invisible. This implies a temporal relationship between the wave-like structure and the burst. In order to confirm this point, we do further analysis and in particular derive the propagation speed of the wave-like structure and that of the type II radio burst.

\subsection{Wave Speed measured from AIA images}

First, we measure the propagation speed of the wave-like structure from the EUV observations of {\it SDO}/AIA at the 171 {\AA}, 193 {\AA} and 211 {\AA} wavelength channels. To do so, we put a slice (shown as the pink dashed line in Figure 3) along the propagation direction of the wave in the running-difference images at these channels. The time-slice diagrams are then plotted in Figure 5, whose time range is from 07:30 to 07:50 UT, covering the whole process of the formation and propagation of the wave-like structure. The wave-like structure can be identified clearly in the time-slice images as a bright narrow stripe. The front of the wave-like structure is marked by green asterisks in the images. The time interval of these asterisks is fixed to be 24 s. Because the wave front is identified visually, to reduce the error, we try in total 10 times to obtain the front points and then calculate the mean value and the standard deviation, which are shown in Figure 5.

The speed of the wave-like structure is shown in the bottom row of Figure 5. It can be seen that the three channels yield similar results. The initial speed of the wave was about 1000 km s$^{-1}$, and then decreased to about 600 km s$^{-1}$ in 1 minute. After that, the wave propagated at a nearly constant speed until its disappearance. The linear speed of the wave is calculated to be 539 km s$^{-1}$, 654 km s$^{-1}$ and 659 km s$^{-1}$ in the 171 {\AA}, 193 {\AA} and 211 {\AA} channels, respectively. Such a speed is marginal to generate a shock as the Alfv\'en wave speed in the corona is very similar to that. Note that in the top row of Figure 5, some bright structures below the wave front refer to the expansion of the loops. The loop expansion in the initiation phase of the wave-like structure is very hard to identify, which could be submerged by the complicated background emission in the active region.

\subsection{Shock Speed Derived from the Radio Spectrum}

It is well known that the frequency drift in the dynamic spectrum for a type II radio burst is produced by a shock propagating outwards in the corona. In order to get the shock speed from the dynamic spectrum, we need to know the density distribution of the corona. However, most of the density models available may deviate significantly from the real case and affect the inversion accuracy of the shock speed. In previous studies, the most commonly used density models are based on the models by \citet{Newkirk1961} and \citet{Saito1977}. Usually, these models are assumed to be multiplied by a coefficient \citep[e.g.,][]{Bemporad2010,Feng2012,Magdalenic2012}. This coefficient is not an invariant, but depends sensitively on time (solar maximum and minimum years) and on place (active region and quiet region). Therefore, it is still a problem how to adopt a coefficient that is accurate enough for a specific event.

Since the start frequency of the type II radio burst is 109 MHz, the start height of the shock is estimated to be less than 0.1R$_{\odot}$ under the \citet{Saito1977} model, which should be within the FOV of AIA. Indeed, the wave-like structure did appear in the FOV of the AIA, as stated above. Thanks to the multi-wavelength EUV emissions observed by AIA, it is possible to derive the density and temperature structure of the coronal part above the active region using the DEM method \citep{Cheng2012}. Then, we can get a more accurate density model, so as to derive a reliable shock speed from the type II burst, which can then be compared with the wave speed measured in the AIA images.

\subsubsection{Constraining the Density Model}

The DEM method is applied to calculate the density maps of the corona above the active region. The procedure is as follows. We use the intensities in six wavelength channels (94 {\AA}, 131 {\AA}, 171, {\AA}, 193 {\AA}, 211 {\AA}, and 335 {\AA}) recorded by {\it SDO}/AIA to derive the emission measure (EM) distribution. We choose the AIA images at 07:20 UT, 20 minutes before the start time of type II radio burst, to get the EM map, which is shown in the left panel of Figure 6.

The density of the emitting plasma can be calculated from the EM as \citep{Cheng2012},
\begin{equation}
    n_{e} = \sqrt{\frac{EM}{s}}, \\
\end{equation}
where $s$ is the line-of-sight length and can be estimated as \citep{Zucca2014},
\begin{equation}
    s \sim \sqrt{H \pi r},  \\
\end{equation}
where $H$ is the pressure scale height and $r$ is the heliocentric distance of the active region. The density distribution derived from the DEM method is then used to construct a specific density model for this event.

We first calculate the density distribution for the coronal part between the wave and the active region below. This part is shown as the region enclosed by a white box in Figure 6, which has an angular range of $24\degree$, and a radial range of 1.02--1.06 R$_{\odot}$. With the density distribution for this region, we then calculate the mean density at a specific height by making an average over the whole angular range. In this way, we can get a radial distribution of the density, which is shown as the black solid line in the right panel of Figure 6. This result is then used to constrain the density model of \citet{Saito1977}. For simplicity, we multiply the density model by a coefficient in order to match the derived density distribution. The optimal coefficient is found to be 0.8. The revised density model is shown as the black dashed line in the right panel of Figure 6. Note that the revised density model can match the derived density distribution almost perfectly with a correlation coefficient of 0.98 between them.

\subsubsection{Shock Speed Derived from the Frequency Drift Rate}

With the appropriate density model, we can deduce the shock speed from the radio dynamic spectrum. First, we can get the frequency as a function of time ($t$) for the type II radio burst from the frequency spectrum. The frequency of type II radio bursts is considered to be the plasma frequency of the local electrons around the shock. Moreover, the plasma frequency is related to the electron density as \citep[e.g.,][]{Gopalswamy2009,Ma2011},
\begin{equation}
    f_{p} = 8.98 \times 10^{3} \sqrt{n_{e}}.\\
\end{equation}
Then, for a specific frequency of the burst, the density at the shock producing the burst is calculated and the corresponding height ($h$) can be found from the revised density distribution. Finally, we can get the speed of the shock from the frequency drift rate.

In practice, we select 15 points from the beginning to the end of the radio burst in the fundamental frequency belt, as marked with red asterisks in the left panel of Figure 7. It is seen that the fundamental belt becomes weak at the later period of the burst, while the harmonic belt is still fairly strong though dispersive. Both of the belts ended at about 07:44 UT. As already mentioned, the start time of the burst is about 120 s later than the appearance of the wave in the AIA images, while the end time of the burst is about 120 s later than the disappearance of the wave in the AIA images. The red solid curves in the left panel are the fitting to the fundamental and harmonic frequency belts. With the frequency-time curve, we can obtain the height-time relation and hence the speed of the shock. The result is shown in the right panels of Figure 7. It is shown that the shock speed is about 700 km s$^{-1}$ at the beginning and then decreases to 450 km s$^{-1}$ later. The average speed is 542 km s$^{-1}$. The speed of the shock is close to the speed of the wave which has been detected in the AIA images.

Considering the similarity in speeds and the close relationship in time between the wave-like structure and the type II radio burst, we believe that the former is indeed a shock that is responsible for the type II radio burst in the dynamic spectrum.

\section{Discussion}
\label{sect:discussion}

According to previous studies, a common view is that the coronal shock, responsible for the type II radio burst, is generated either by CMEs or flares, although more people believe the CME scenario. For this event, we checked the coronal images of LASCO/C2 and {\it STEREO}/COR1 (A and B) in white light and {\it SDO}/AIA in EUV, and did not find occurrence of any associated CME. If a CME was erupting, it should be detectable since the source region is close to the solar limb. Therefore, we can conclude that the shock is not likely driven by a CME as usual. On the other hand, it was sometimes argued that the pressure pulse in a solar flare might generate a blast shock wave, which can emit type II radio bursts \citep[e.g.,][]{Uchida1974}. Similarly to the discussions in \citet{Vrsnak2008}, we tend to ignore this possibility because the radio burst event in this paper is associated with a C2.4-flare, whereas quite some X-class flares, which are $\sim$100 times stronger than C-class flare, are not associated with EUV waves or type II radio bursts, e.g., the 2005 January 14 flare studied by \citet{Chen2006}.

Considering the above facts, here we are trying to propose a new mechanism for the generation of the shock wave in the case of a CME-less flare. By inspecting carefully the AIA images, we find that some coronal loops under the shock wave front expanded obviously, starting from the appearance of the shock wave and ending a few minutes after the disappearance of the wave. We think that such a coincidence between the loop expansion and the coronal shock is not incidental, but implying a casual relationship. To check this point, we further analyze the HMI magnetograms of the active region and find that there were continual magnetic emergence and magnetic cancellation near the eastern feet of the loops (see Figure 8). It has been well established that magnetic reconnection between an emerging magnetic flux and the pre-existing coronal loop would lead to an impulsive flare \citep{Heyvaerts1977}. At the same time, such a reconnection would also lead to the expansion of the pre-existing coronal loop after it is re-rooted to another site due to the interchange reconnection, as simulated by \citet{Chen2000}.

The post-reconnection coronal loop, as indicated by the dashed line in the left panel of Figure 9, is kinked around the reconnection site. Such a kink results in a strong magnetic tension force along the direction of the reconnection jet, which drives the coronal loop to expand. Such an expanding loop would lead to a fast-mode magnetoacoustic wave propagating outward, which further steepens to form a shock wave due to the nonlinear effect, since the background fast-mode wave speed decreases with height. The shock is then manifested as the type II radio burst. Since there is no continual eruption like CMEs, the shock exists for a short time and then decays to an ordinary wave, which can explain why the radio burst lasted for a short period as indicated by Figure 1.

However, it should be mentioned that this type of interchange reconnection happens frequently as long as newly emerging magnetic flux meets with the pre-existing coronal loop, whereas shock waves responsible for type II radio bursts are rare in these CME-less flare events. This implies that a special condition is required. Here, we conjecture two different scenarios: (1) if the pre-existing magnetic loop is strongly inclined near the emerging flux, as depicted in the left panel of Figure 9, there would be a shock wave generated since the magnetic loop after reconnection (dashed line) is strongly bent. The magnetic tension force is very strong in this case. With the strong magnetic tension force, a faster jet would be ejected, which takes the threading magnetic field lines away quickly, and  a fast-mode shock wave is expected to be produced. As a strongly-bent field lines stretch, they become less bent, and no continual pushing is provided for the shock wave. As the consequence, the shock wave decays rapidly as it propagates outward, which naturally explains the short lifetime of the type II radio burst as indicated by Figure 1; (2) if the pre-existing magnetic loop is slightly inclined near the emerging flux, as depicted in the right panel of Figure 9, there would be no shock wave generated, and only a fast-mode MHD wave is generated since the magnetic loop after reconnection (dashed line) is weakly bent. The magnetic tension force is weak in this case. In our event, the magnetic field lines as traced from the EUV loops, which are indicated by the blue arrow in Figure 4, are strongly bent toward the solar surface. This fits the first scenario and explains why a type II radio burst is observed.

In order to further confirm such a conjecture, we perform two-dimensional magnetohydrodynamic numerical simulations of the interchange reconnection between emerging flux and pre-existing magnetic field in two cases. In case A, the magnetic field lines are strongly inclined near the flux emerging site, as illustrated in the top-left panel of Figure 10, whereas in case B, the pre-existing magnetic field lines are slightly inclined near the flux emerging site, as illustrated in the top-right panel of Figure 10. Note that the magnetic field strength near the reconnection is chosen to be the same. We adopt the same numerical code as in \citet{Chen2000}, which employs a multi-step implicit scheme \citep{Hu1989, Chen2000Ch}. The numerical results reveal that as magnetic flux emerges from the bottom of the simulation box, i.e., the base of the corona, interchange reconnection occurs between the emerging magnetic flux and the pre-existing field lines.  In both cases, the bent field lines after reconnection drive a jet and a wave front propagating in the upper-right direction. The wave fronts are indicated in the two panels of the bottom row of Figure 10 for both cases. However, it is noted that in case A, the density enhancement of the wave front is stronger than that in case B, which is manifested by different colors in the density map. Such a result strongly implies that only with strongly bent post-reconnection field lines can a shock wave be produced when emerging flux reconnects with the pre-existing coronal field in CME-less flares.

\section{Conclusions}
\label{sect:conclusion}
In this paper, we study a type II radio burst without an accompanying CME. The burst is related to a coronal shock that can be clearly identified in AIA images. However, we find that the shock is neither produced by a CME nor by a flare. We propose a new mechanism for this event: the shock is generated by the expansion of the magnetic loops after reconnecting with emerging flux. The main results of the paper are summarized as follows:

1. We apply the DEM method to the multi-wavelength EUV data to obtain the density distribution in the radial direction above the active region, which is then used to constrain the density model of \citet{Saito1977}. Using the resulting density model and the dynamic frequency spectrum of the type II radio burst, we further derive the shock speed.

2. We identify the wave-like structure in the AIA images. We find a close relationship in the occurrence time and in the propagation speeds between the wave-like structure and the shock inverted from the type II radio burst. The wave-like structure in AIA images is thus most likely a coronal shock responsible for the type II burst.

3. We suggest that the shock is neither generated by a CME as a piston-driven wave nor by a flare as a blast wave. We propose a new mechanism for the generation of the shock. The strongly-inclined magnetic loops, after reconnecting with emerging flux, can expand quickly to generate a shock front ahead. This scenario of loop-driven shock is somewhat different from the usual mechanism of CME-driven shock. In the former, the expansion of loops serves as a piston driving the shock wave in a short time while the loops do not erupt as a CME with their feet still anchored to the photosphere.

\acknowledgements

The authors are grateful to the referee for valuable comments that helped improve the manuscript. SDO is a mission of NASA's Living With a Star Program, STEREO is the third mission in NASA's Solar Terrestrial Probes program, and SOHO is a mission of international cooperation between ESA and NASA. This work was supported by the NSFC (grants 11303016, 11373023, 11203014, and 11025314) and NKBRSF (grants 2011CB811402 and 2014CB744203).


\bibliographystyle{apj}
\bibliography{reference}

\begin{thebibliography}{41}
\expandafter\ifx\csname natexlab\endcsname\relax\def\natexlab#1{#1}\fi

\bibitem[{{Bemporad} \& {Mancuso}(2010)}]{Bemporad2010}
{Bemporad}, A., \& {Mancuso}, S. 2010, \apj, 720, 130

\bibitem[{{Brueckner} {et~al.}(1995){Brueckner}, {Howard}, {Koomen},
  {Korendyke}, {Michels}, {Moses}, {Socker}, {Dere}, {Lamy}, {Llebaria},
  {Bout}, {Schwenn}, {Simnett}, {Bedford}, \& {Eyles}}]{Brueckner1995}
{Brueckner}, G.~E., {et~al.} 1995, \solphys, 162, 357

\bibitem[{{Cane} {et~al.}(1987){Cane}, {Sheeley}, \& {Howard}}]{Cane1987}
{Cane}, H.~V., {Sheeley}, Jr., N.~R., \& {Howard}, R.~A. 1987, \jgr, 92, 9869

\bibitem[{{Chen} {et~al.}(2000){Chen}, {Fang}, \& {Hu}}]{Chen2000Ch}
{Chen}, P., {Fang}, C., \& {Hu}, Y. 2000, Chinese Science Bulletin, 45, 798

\bibitem[{{Chen}(2006)}]{Chen2006}
{Chen}, P.~F. 2006, \apjl, 641, L153

\bibitem[{{Chen}(2011)}]{Chen2011}
---. 2011, Living Reviews in Solar Physics, 8, 1

\bibitem[{{Chen} \& {Shibata}(2000)}]{Chen2000}
{Chen}, P.~F., \& {Shibata}, K. 2000, \apj, 545, 524

\bibitem[{{Cheng} {et~al.}(2012){Cheng}, {Zhang}, {Saar}, \&
  {Ding}}]{Cheng2012}
{Cheng}, X., {Zhang}, J., {Saar}, S.~H., \& {Ding}, M.~D. 2012, \apj, 761, 62

\bibitem[{{Cho} {et~al.}(2013){Cho}, {Gopalswamy}, {Kwon}, {Kim}, \&
  {Yashiro}}]{Cho2013}
{Cho}, K.-S., {Gopalswamy}, N., {Kwon}, R.-Y., {Kim}, R.-S., \& {Yashiro}, S.
  2013, \apj, 765, 148

\bibitem[{{Cliver} {et~al.}(2004){Cliver}, {Nitta}, {Thompson}, \&
  {Zhang}}]{Cliver2004}
{Cliver}, E.~W., {Nitta}, N.~V., {Thompson}, B.~J., \& {Zhang}, J. 2004,
  \solphys, 225, 105

\bibitem[{{Feng} {et~al.}(2012){Feng}, {Chen}, {Kong}, {Li}, {Song}, {Feng}, \&
  {Liu}}]{Feng2012}
{Feng}, S.~W., {Chen}, Y., {Kong}, X.~L., {Li}, G., {Song}, H.~Q., {Feng},
  X.~S., \& {Liu}, Y. 2012, \apj, 753, 21

\bibitem[{{Giacalone}(2005)}]{Giacalone2005}
{Giacalone}, J. 2005, \apjl, 628, L37

\bibitem[{{Gopalswamy}(2001)}]{Gopalswamy2001}
{Gopalswamy}, N. 2001, \jgr, 106, 25135

\bibitem[{{Gopalswamy} {et~al.}(2009){Gopalswamy}, {Thompson}, {Davila},
  {Kaiser}, {Yashiro}, {M{\"a}kel{\"a}}, {Michalek}, {Bougeret}, \&
  {Howard}}]{Gopalswamy2009}
{Gopalswamy}, N., {et~al.} 2009, \solphys, 259, 227

\bibitem[{{Heyvaerts} {et~al.}(1977){Heyvaerts}, {Priest}, \&
  {Rust}}]{Heyvaerts1977}
{Heyvaerts}, J., {Priest}, E.~R., \& {Rust}, D.~M. 1977, \apj, 216, 123

\bibitem[{{Howard} {et~al.}(2008){Howard}, {Moses}, {Vourlidas}, {Newmark},
  {Socker}, {Plunkett}, {Korendyke}, {Cook}, {Hurley}, {Davila}, {Thompson},
  {St Cyr}, {Mentzell}, {Mehalick}, {Lemen}, {Wuelser}, {Duncan}, {Tarbell},
  {Wolfson}, {Moore}, {Harrison}, {Waltham}, {Lang}, {Davis}, {Eyles},
  {Mapson-Menard}, {Simnett}, {Halain}, {Defise}, {Mazy}, {Rochus}, {Mercier},
  {Ravet}, {Delmotte}, {Auchere}, {Delaboudiniere}, {Bothmer}, {Deutsch},
  {Wang}, {Rich}, {Cooper}, {Stephens}, {Maahs}, {Baugh}, {McMullin}, \&
  {Carter}}]{Howard2008}
{Howard}, R.~A., {et~al.} 2008, \ssr, 136, 67

\bibitem[{Hu(1989)}]{Hu1989}
Hu, Y. 1989, Journal of Computational Physics, 84, 441

\bibitem[{{Kaiser} {et~al.}(2008){Kaiser}, {Kucera}, {Davila}, {St.~Cyr},
  {Guhathakurta}, \& {Christian}}]{Kaiser2008}
{Kaiser}, M.~L., {Kucera}, T.~A., {Davila}, J.~M., {St.~Cyr}, O.~C.,
  {Guhathakurta}, M., \& {Christian}, E. 2008, \ssr, 136, 5

\bibitem[{{Kouloumvakos} {et~al.}(2013){Kouloumvakos}, {Patsourakos},
  {Hillaris}, {Vourlidas}, {Preka-Papadema}, {Moussas}, {Caroubalos},
  {Tsitsipis}, \& {Kontogeorgos}}]{Kouloumvakos2013}
{Kouloumvakos}, A., {et~al.} 2013, ArXiv e-prints

\bibitem[{{Leblanc} {et~al.}(2001){Leblanc}, {Dulk}, {Vourlidas}, \&
  {Bougeret}}]{Leblanc2001}
{Leblanc}, Y., {Dulk}, G.~A., {Vourlidas}, A., \& {Bougeret}, J.-L. 2001, \jgr,
  106, 25301

\bibitem[{{Lemen} {et~al.}(2012){Lemen}, {Title}, {Akin}, {Boerner}, {Chou},
  {Drake}, {Duncan}, {Edwards}, {Friedlaender}, {Heyman}, {Hurlburt}, {Katz},
  {Kushner}, {Levay}, {Lindgren}, {Mathur}, {McFeaters}, {Mitchell}, {Rehse},
  {Schrijver}, {Springer}, {Stern}, {Tarbell}, {Wuelser}, {Wolfson}, {Yanari},
  {Bookbinder}, {Cheimets}, {Caldwell}, {Deluca}, {Gates}, {Golub}, {Park},
  {Podgorski}, {Bush}, {Scherrer}, {Gummin}, {Smith}, {Auker}, {Jerram},
  {Pool}, {Soufli}, {Windt}, {Beardsley}, {Clapp}, {Lang}, \&
  {Waltham}}]{Lemen2012}
{Lemen}, J.~R., {et~al.} 2012, \solphys, 275, 17

\bibitem[{{Liu} {et~al.}(2009){Liu}, {Luhmann}, {Bale}, \& {Lin}}]{Liu2009}
{Liu}, Y., {Luhmann}, J.~G., {Bale}, S.~D., \& {Lin}, R.~P. 2009, \apjl, 691,
  L151

\bibitem[{{Ma} {et~al.}(2011){Ma}, {Raymond}, {Golub}, {Lin}, {Chen}, {Grigis},
  {Testa}, \& {Long}}]{Ma2011}
{Ma}, S., {Raymond}, J.~C., {Golub}, L., {Lin}, J., {Chen}, H., {Grigis}, P.,
  {Testa}, P., \& {Long}, D. 2011, \apj, 738, 160

\bibitem[{{Magdaleni{\'c}} {et~al.}(2012){Magdaleni{\'c}}, {Marqu{\'e}},
  {Zhukov}, {Vr{\v s}nak}, \& {Veronig}}]{Magdalenic2012}
{Magdaleni{\'c}}, J., {Marqu{\'e}}, C., {Zhukov}, A.~N., {Vr{\v s}nak}, B., \&
  {Veronig}, A. 2012, \apj, 746, 152

\bibitem[{{Magdaleni{\'c}} {et~al.}(2008){Magdaleni{\'c}}, {Vr{\v s}nak},
  {Pohjolainen}, {Temmer}, {Aurass}, \& {Lehtinen}}]{Magdalenic2008}
{Magdaleni{\'c}}, J., {Vr{\v s}nak}, B., {Pohjolainen}, S., {Temmer}, M.,
  {Aurass}, H., \& {Lehtinen}, N.~J. 2008, \solphys, 253, 305

\bibitem[{{Mann} \& {Classen}(1995)}]{Mann1995}
{Mann}, G., \& {Classen}, H.-T. 1995, \aap, 304, 576

\bibitem[{{Miteva} \& {Mann}(2007)}]{Miteva2007}
{Miteva}, R., \& {Mann}, G. 2007, \aap, 474, 617

\bibitem[{{Newkirk}(1961)}]{Newkirk1961}
{Newkirk}, Jr., G. 1961, \apj, 133, 983

\bibitem[{{Nindos} {et~al.}(2011){Nindos}, {Alissandrakis}, {Hillaris}, \&
  {Preka-Papadema}}]{Nindos2011}
{Nindos}, A., {Alissandrakis}, C.~E., {Hillaris}, A., \& {Preka-Papadema}, P.
  2011, \aap, 531, A31

\bibitem[{{Pesnell} {et~al.}(2012){Pesnell}, {Thompson}, \&
  {Chamberlin}}]{Pesnell2012}
{Pesnell}, W.~D., {Thompson}, B.~J., \& {Chamberlin}, P.~C. 2012, \solphys,
  275, 3

\bibitem[{{Reiner} {et~al.}(2001{\natexlab{a}}){Reiner}, {Kaiser}, \&
  {Bougeret}}]{Reiner2001}
{Reiner}, M.~J., {Kaiser}, M.~L., \& {Bougeret}, J.-L. 2001{\natexlab{a}},
  \jgr, 106, 29989

\bibitem[{{Reiner} {et~al.}(2001{\natexlab{b}}){Reiner}, {Kaiser},
  {Gopalswamy}, {Aurass}, {Mann}, {Vourlidas}, \&
  {Maksimovic}}]{Reiner2001stat}
{Reiner}, M.~J., {Kaiser}, M.~L., {Gopalswamy}, N., {Aurass}, H., {Mann}, G.,
  {Vourlidas}, A., \& {Maksimovic}, M. 2001{\natexlab{b}}, \jgr, 106, 25279

\bibitem[{{Saito} {et~al.}(1977){Saito}, {Poland}, \& {Munro}}]{Saito1977}
{Saito}, K., {Poland}, A.~I., \& {Munro}, R.~H. 1977, \solphys, 55, 121

\bibitem[{{Schwartz} {et~al.}(2011){Schwartz}, {Henley}, {Mitchell}, \&
  {Krasnoselskikh}}]{Schwartz2011}
{Schwartz}, S.~J., {Henley}, E., {Mitchell}, J., \& {Krasnoselskikh}, V. 2011,
  Physical Review Letters, 107, 215002

\bibitem[{{Sckopke}(1995)}]{Sckopke1995}
{Sckopke}, N. 1995, Advances in Space Research, 15, 261

\bibitem[{{Thomsen} {et~al.}(1985){Thomsen}, {Gosling}, {Bame}, \&
  {Mellott}}]{Thomsen1985}
{Thomsen}, M.~F., {Gosling}, J.~T., {Bame}, S.~J., \& {Mellott}, M.~M. 1985,
  \jgr, 90, 137

\bibitem[{{Uchida}(1974)}]{Uchida1974}
{Uchida}, Y. 1974, \solphys, 39, 431

\bibitem[{{Vr{\v s}nak} \& {Cliver}(2008)}]{Vrsnak2008}
{Vr{\v s}nak}, B., \& {Cliver}, E.~W. 2008, \solphys, 253, 215

\bibitem[{{Wild}(1950)}]{Wild1950}
{Wild}, J.~P. 1950, Australian Journal of Scientific Research A Physical
  Sciences, 3, 541

\bibitem[{{Zheleznyakov}(1970)}]{Zheleznyakov1970}
{Zheleznyakov}, V.~V. 1970, {Radio emission of the sun and planets}

\bibitem[{{Zucca} {et~al.}(2014){Zucca}, {Carley}, {Bloomfield}, \&
  {Gallagher}}]{Zucca2014}
{Zucca}, P., {Carley}, E.~P., {Bloomfield}, D.~S., \& {Gallagher}, P.~T. 2014,
  \aap, 564, A47

\end{thebibliography}

\begin{figure}
   \center
   \includegraphics[width=12cm, angle=0]{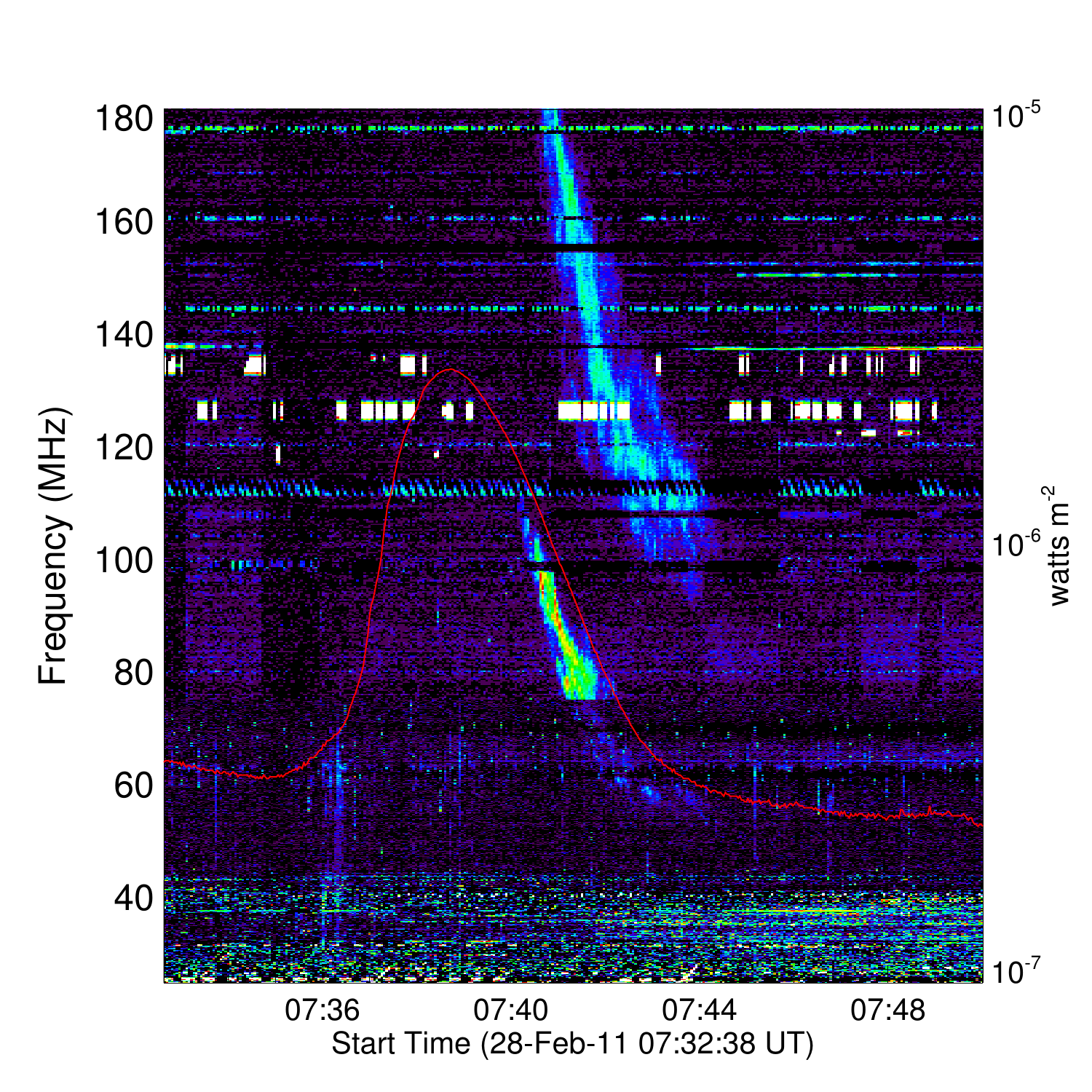}
   \caption{Dynamic spectrogram of the type II radio burst on 2011 February 28 in the range of 25-180 MHz observed by the Learmonth Observatory. The red curve refers to the soft X-ray flux during the burst.}
   \label{spectrum-goes}
\end{figure}
\begin{figure}
   \center
   \includegraphics[width=15cm, angle=0,trim = 0 120 0 -120,clip = true]{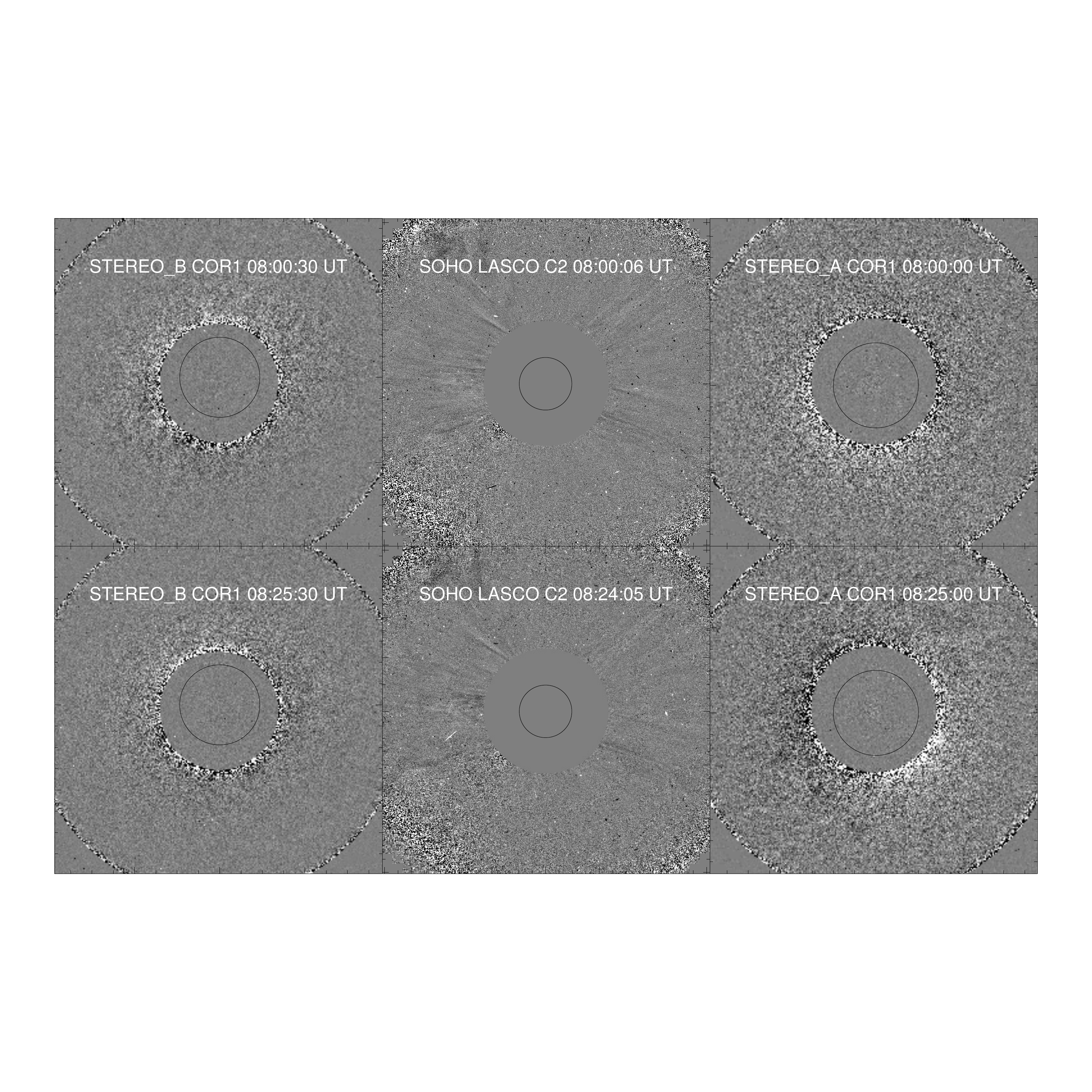}
   \caption{Running difference images of the corona in white-light for two different times after the burst recorded by STEREO-B/COR1 (left), SOHO/C2 (middle) and STEREO-A/COR1 (right).}
   \label{soho-aia-stereo}
\end{figure}
\begin{figure}
   \center
   \includegraphics[width=12cm, angle=0, trim = 0 150 0 150]{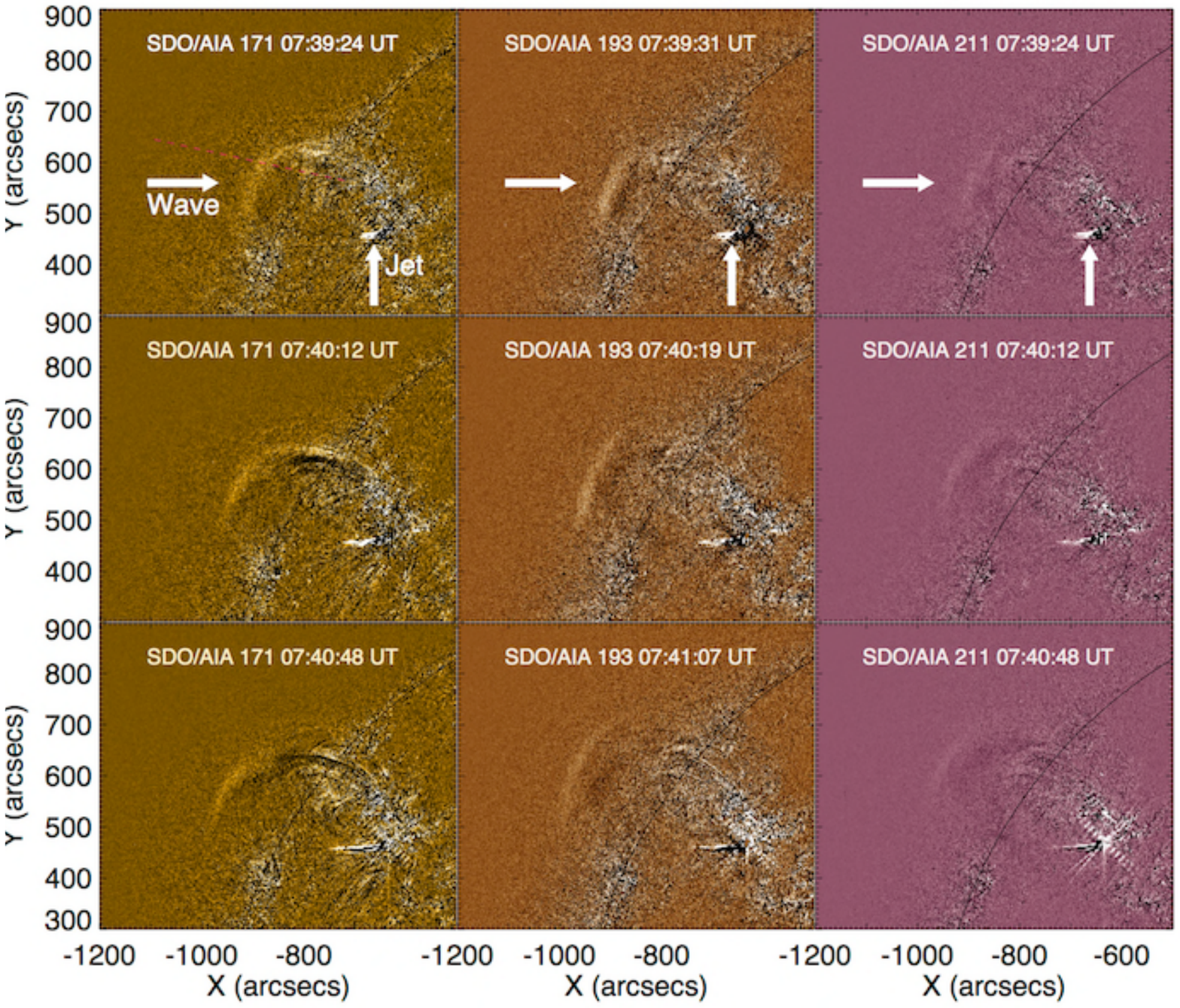}
   \caption{Time sequence of SDO/AIA 171 {\AA} (top), 191 {\AA} (middle), and 211 {\AA} (bottom) running difference images on 2011 February 28. The horizontal arrows indicate the location of the shock wave, while the vertical arrows indicate the jet. The pink dashed line refers to the slice used to trace the wave propagation.}
   \label{profile-wave}
\end{figure}
\begin{figure}
   \center
   \includegraphics[width=18cm, trim = 0 50 0 50,clip = true]{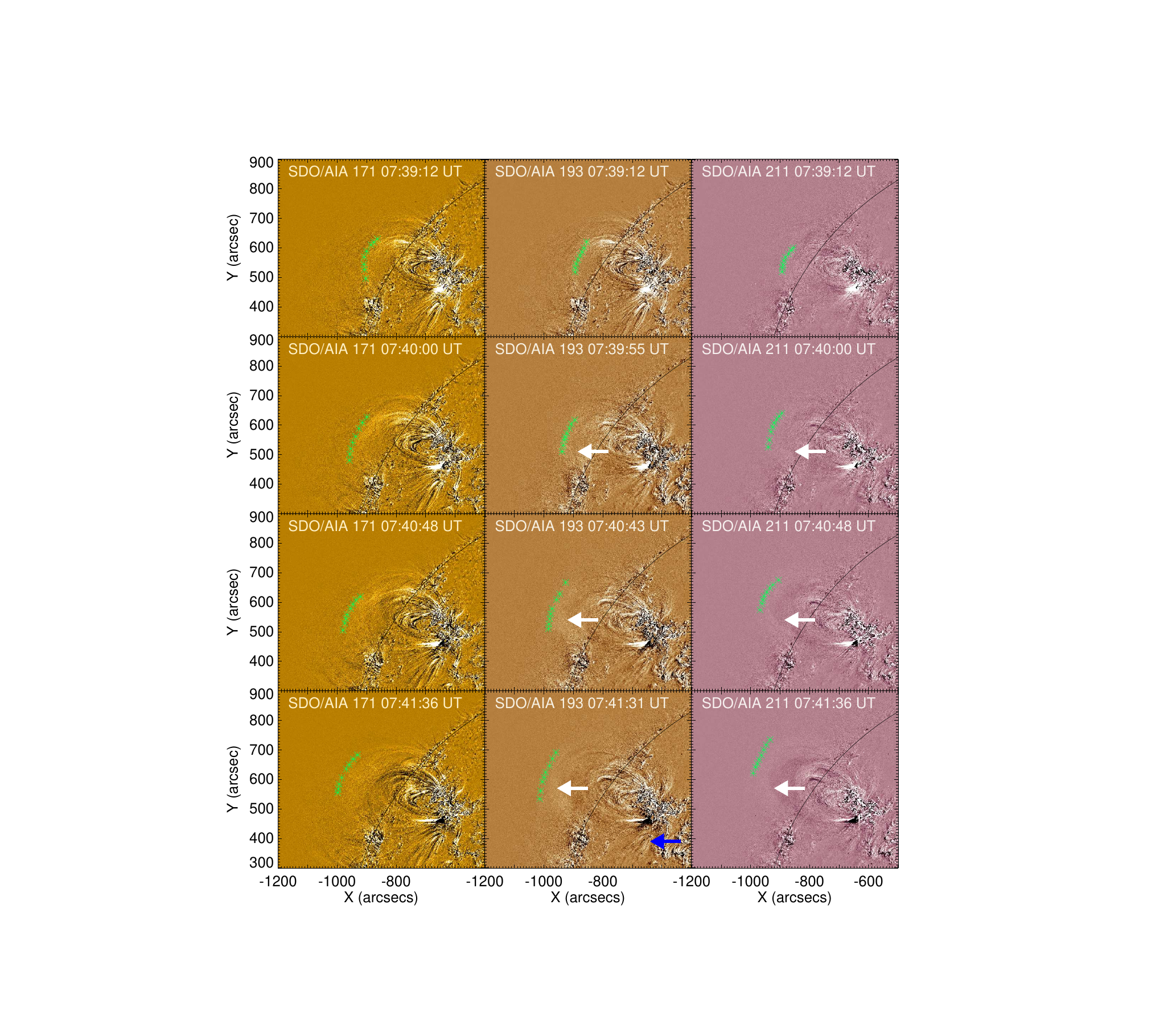}
   \caption{Time sequence of SDO/AIA 171 {\AA} (top), 191 {\AA} (middle), and 211 {\AA} (bottom) running difference images on 2011 February 28. The white arrows indicate the location of the expanding loops. The blue arrow indicates the strongly inclined coronal loops. The green asterisks mark the wave-like structure.}
   \label{profile-loop}
\end{figure}

\begin{figure}
  \center
   \includegraphics[width=16cm, angle=0]{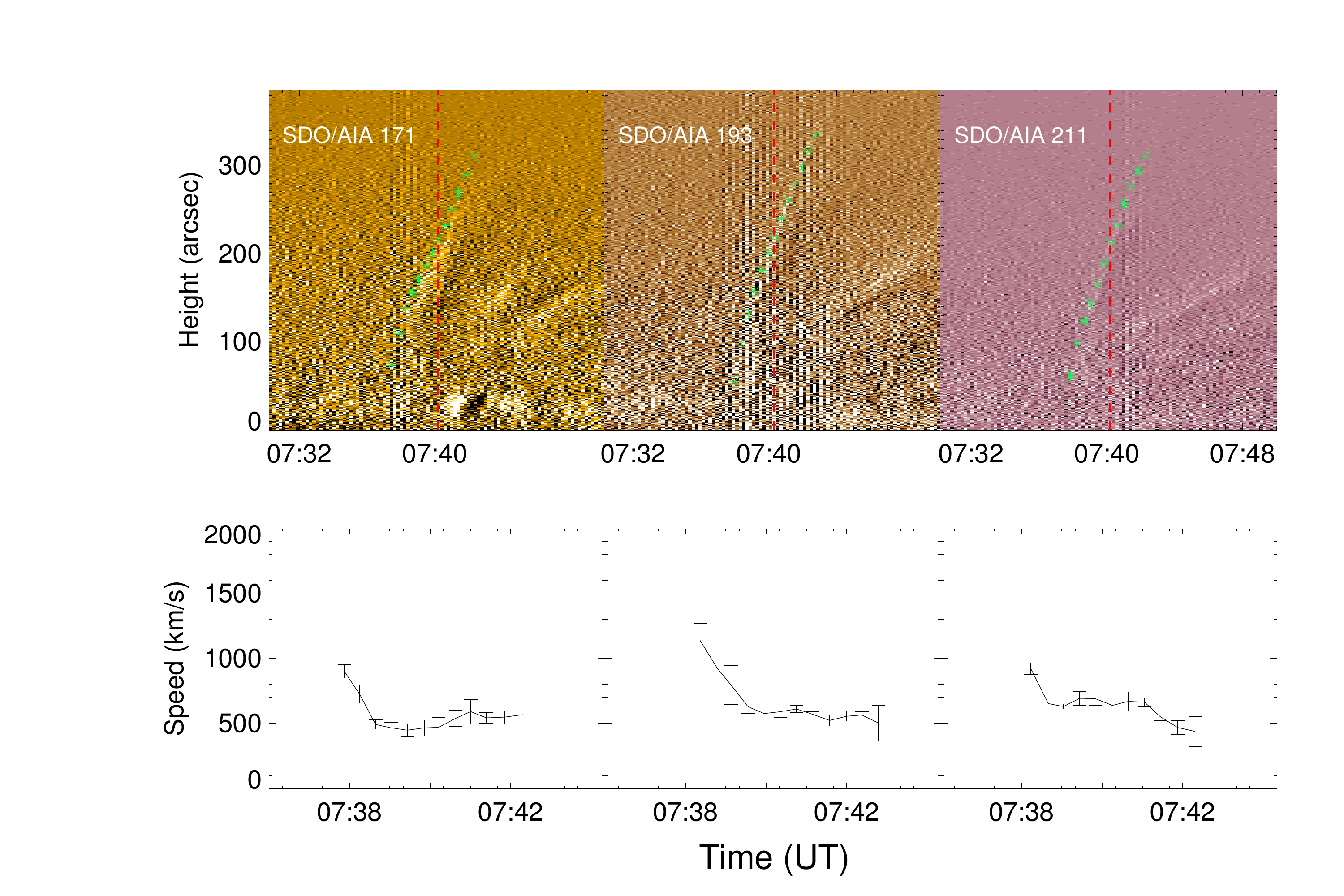}
   \caption{Top row: time-slice plot of the AIA running difference images in 171 {\AA} (left), 193 {\AA} (middle) and 211 {\AA} (right) to show the wave propagation (green asterisks); the red dashed line represent the start time of the type II radio burst. Bottom row: speed of the wave measured in the top panels.}
   \label{vslice-mult}
\end{figure}

\begin{figure}
   \center
   \includegraphics[width=15cm, angle=0]{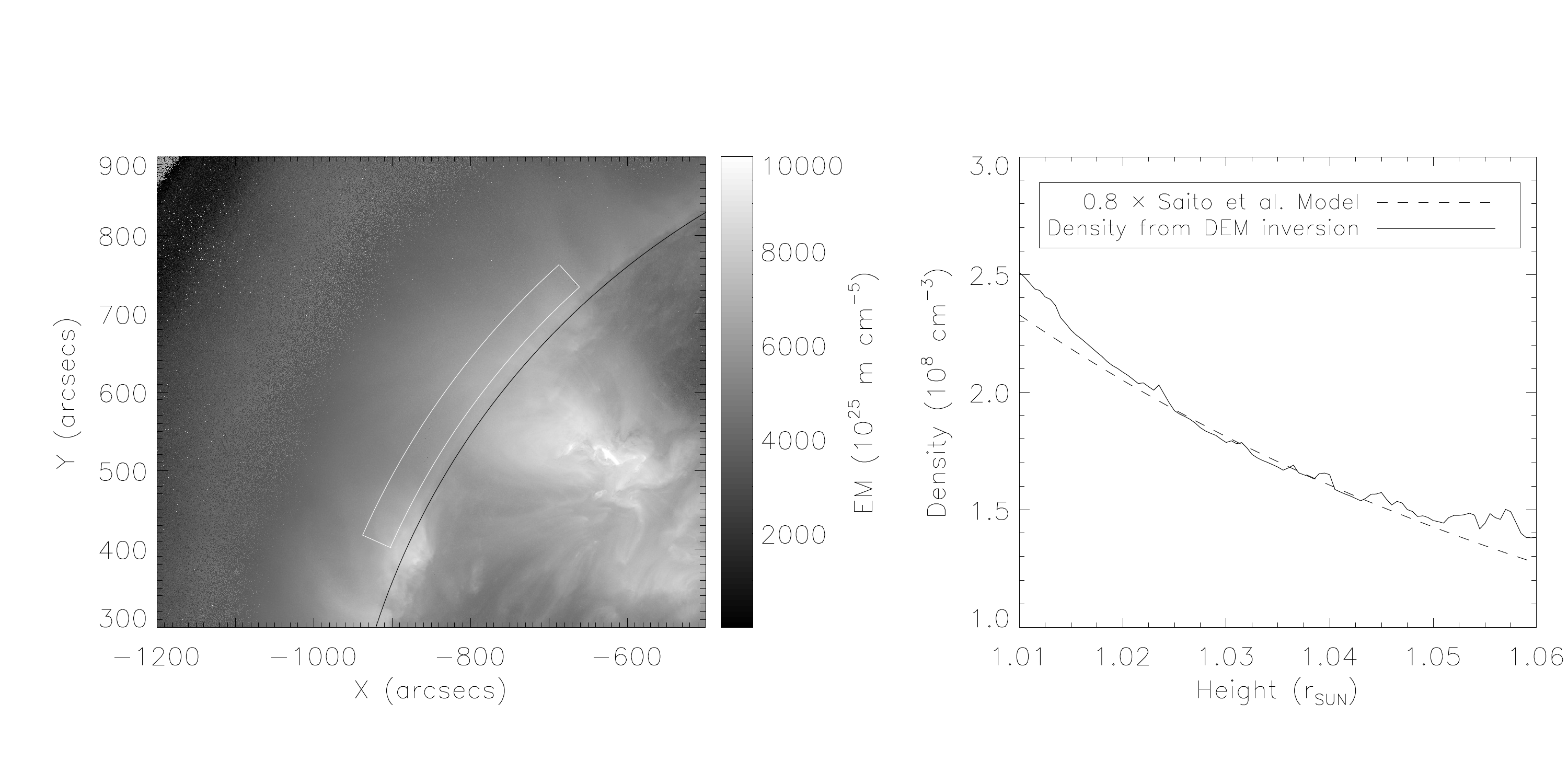}
   \caption{Left: EM map of the active region. Right: density distribution with height. The solid line refers to the radial density distribution of the region encircled by the white line in the left panel, while the dashed line represents the density distribution of the model of Saito et al. (1977) multiplied by a coefficient of 0.8.}
   \label{dem-den}
\end{figure}

\begin{figure}
  \center
   \includegraphics[width=15cm, angle=0]{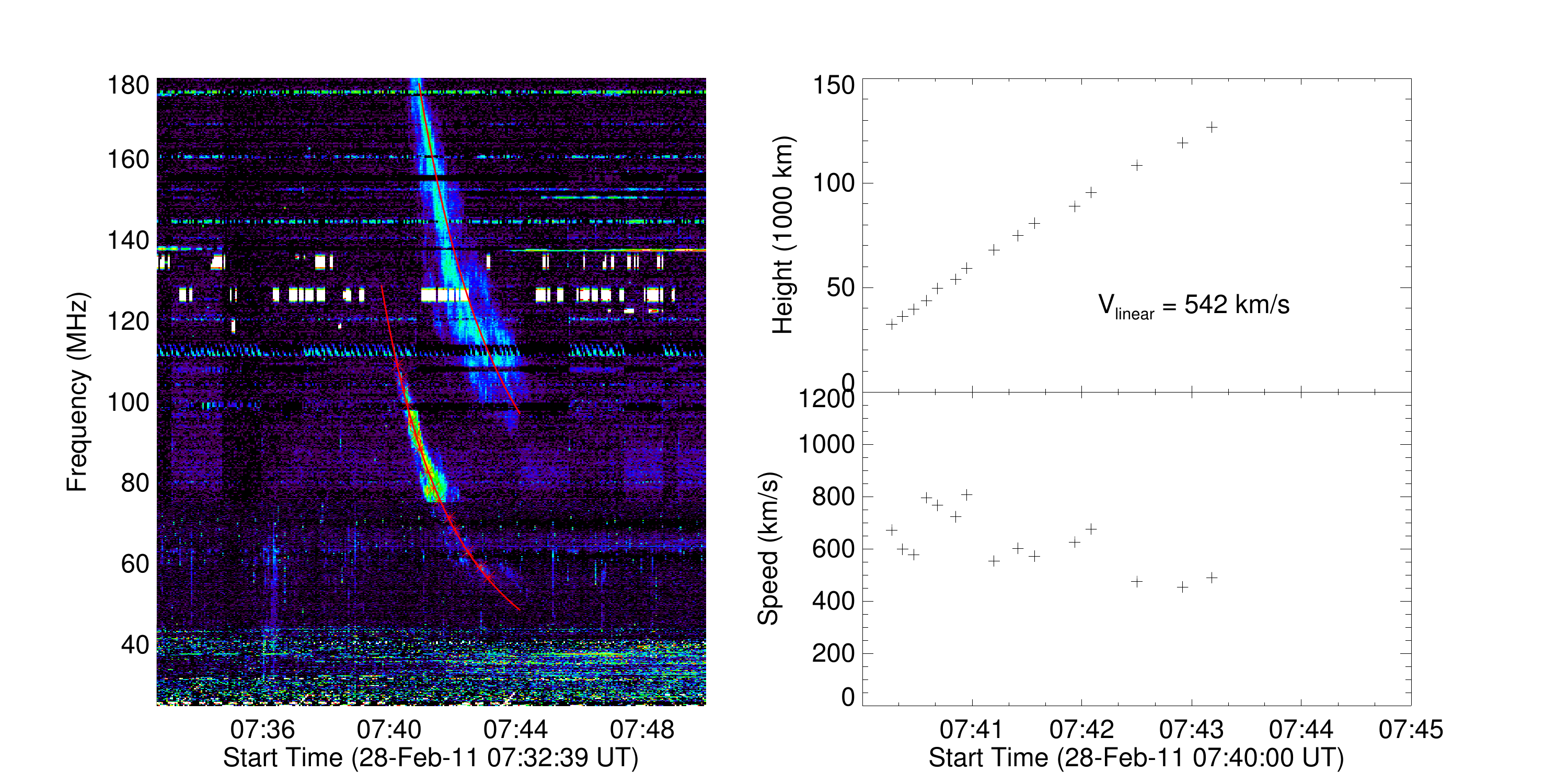}
   \caption{Left: dynamic spectrogram during the type II radio burst. The two red solid curves are the fitting to the frequency-time evolutions for the fundamental and harmonic frequency belts. Right: height and speed of the shock inverted from the revised density model.}
   \label{v-typeii-saito0.800000}
\end{figure}

\begin{figure}
  \center
   \includegraphics[width=15cm, angle=0, trim = 0 150 0 150]{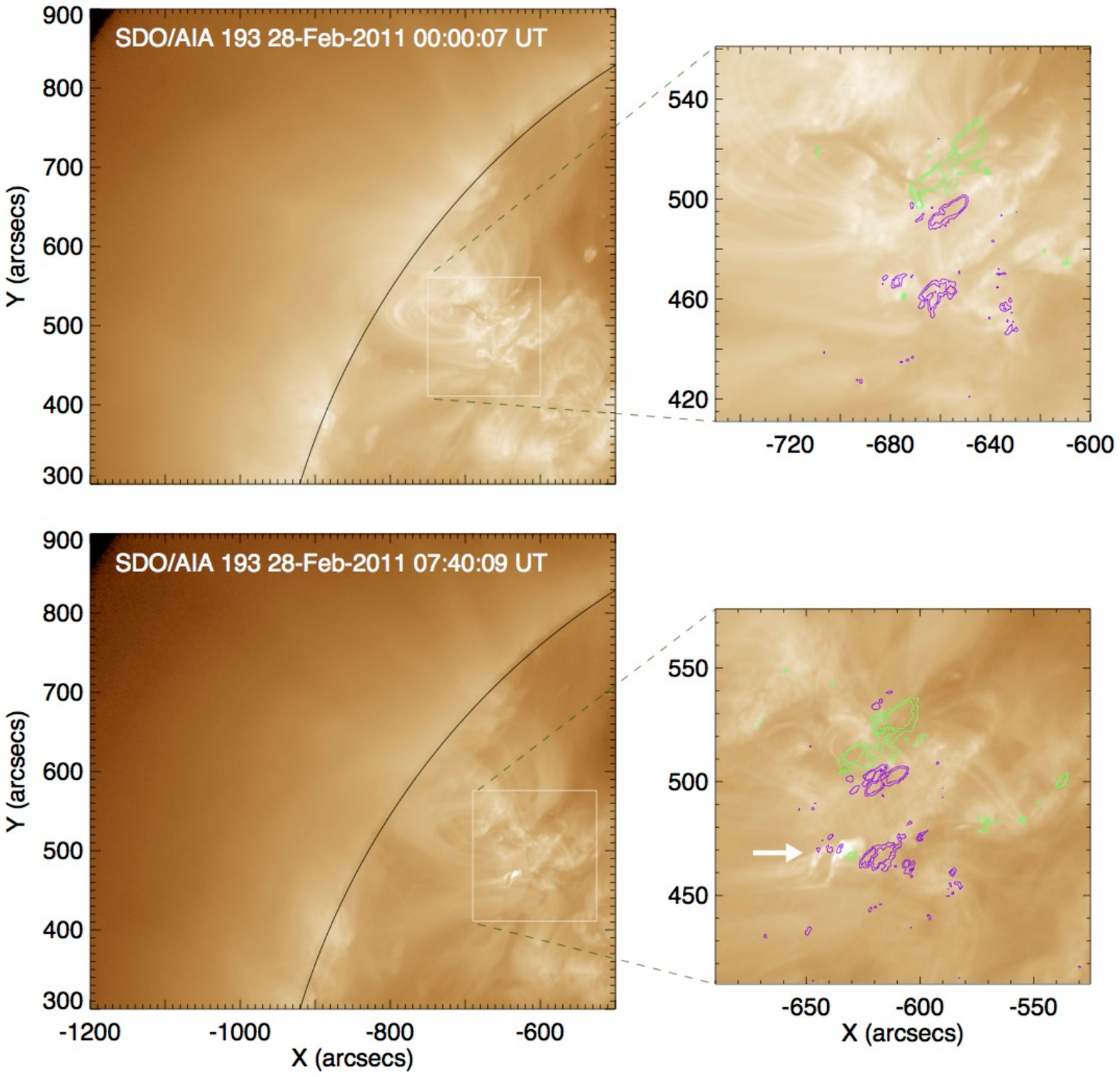}
   \caption{AIA images in 193 {\AA} and HMI magnetograms for a selected area in the active region. Top and bottom panels are for times before and during the event, respectively. The green contours represent negative magnetic polarity, while the purple contours represent the positive polarity.}
   \label{aia-hmi-0228-0000}
\end{figure}

\begin{figure}
    \center
   \includegraphics[width=12cm, angle=0,trim = 0 120 0 120,clip = true]{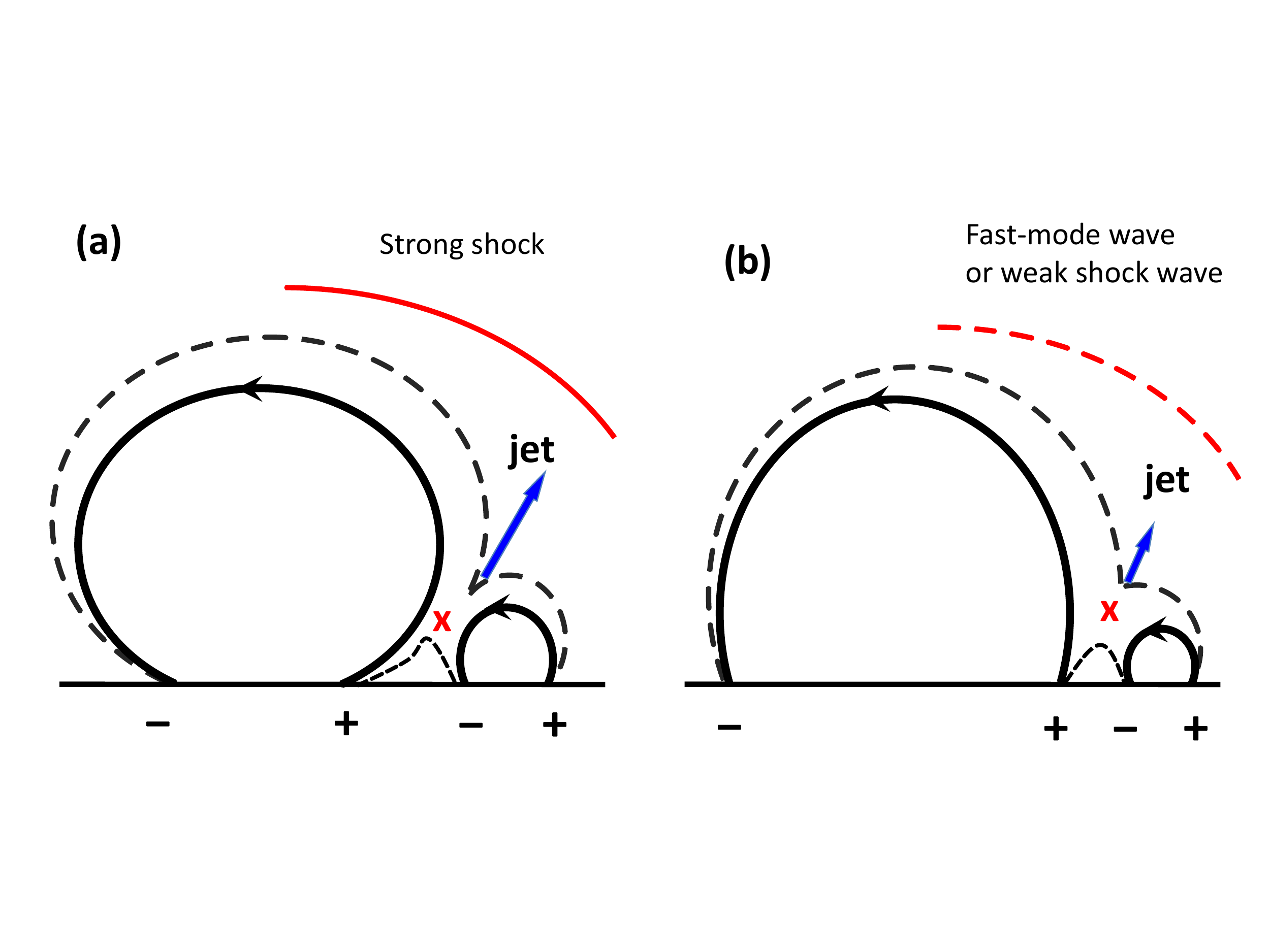}
   \caption{Schematic sketch of the wave generation during interchanged magnetic reconnection. The black solid lines correspond to the magnetic field before reconnection, and the black dashed lines to the magnetic field after reconnection, while the red ``X" represents the reconnection site. The blue arrows indicate the reconnection jets. When the pre-existing magnetic field is strongly inclined as shown in panel (a), the drastic expansion of the strongly kinked post-reconnection loop would result in a strong shock wave as indicated by the red solid line. A type II radio burst is expected in this case. When the pre-existing magnetic field is slightly inclined as shown in panel (b), the slower expansion of the reconnected loop would result in an ordinary MHD wave or a weak shock wave as indicated by the red dashed line. No type II radio burst happens in this case.}
   \label{cartoon}
\end{figure}

\begin{figure}
    \center
   \includegraphics[width=18cm, angle=0]{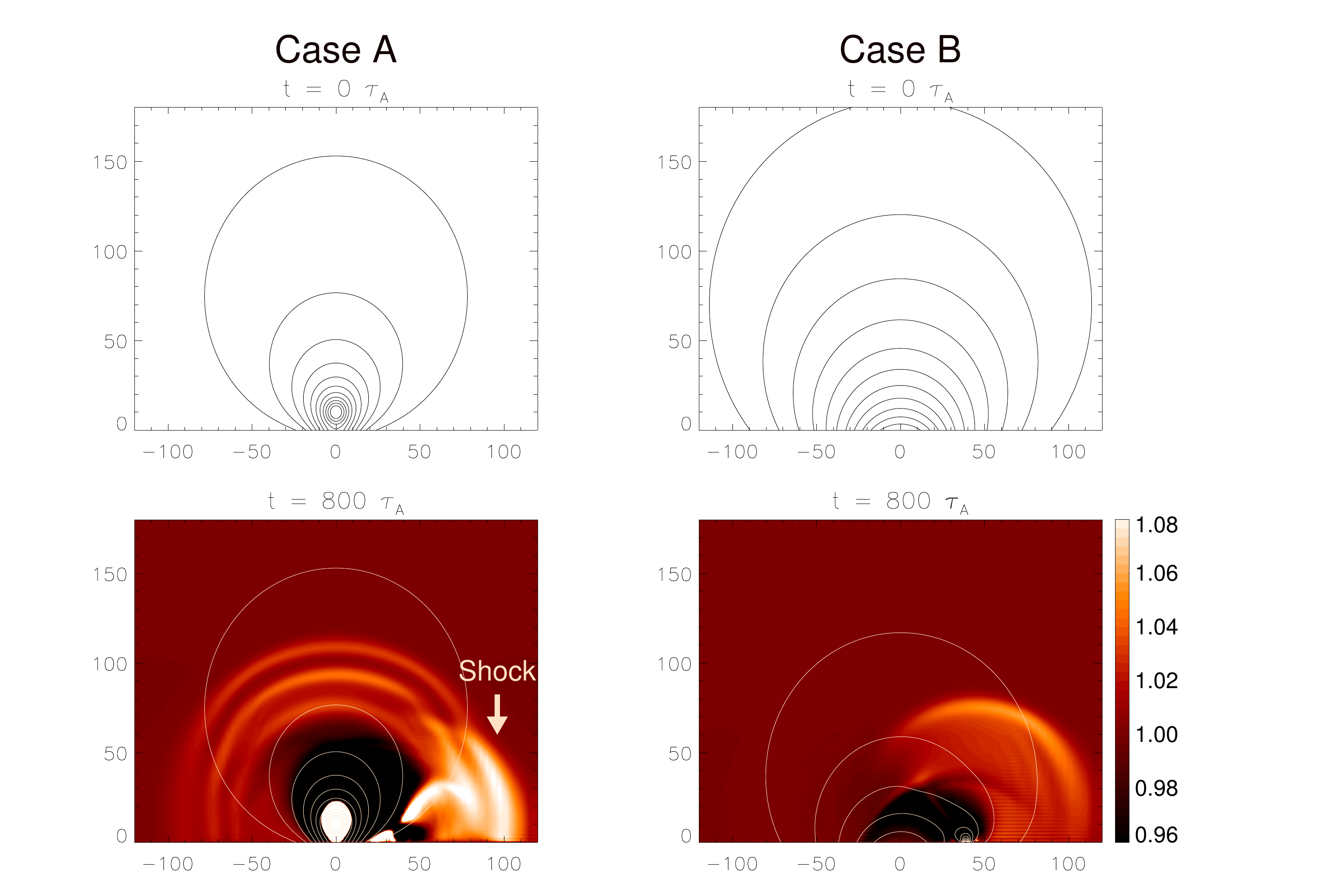}
   \caption{Magnetic field and density evolutions from our MHD numerical simulations. The solid lines correspond to the magnetic field and the color scale to the density. Case A ({\it left}): a strongly-bent post-reconnection loop is produced. The white arrow indicates a shock front. Case B ({\it right}): a weakly-bent post-reconnection loop is produced. Comparing the two cases, it is seen that a much stronger shock wave is produced in case A.}
   \label{simulation}
\end{figure}

\clearpage



\end{document}